\documentclass[]{article}

\usepackage{cite}
\usepackage{graphicx}
\usepackage{amsfonts, amsmath}
\usepackage{tabularx}
\usepackage{booktabs}
\usepackage{url}

\usepackage{comment}
\usepackage{authblk}

\usepackage{xcolor}
\usepackage{color}

\title{The drivers of online polarization: fitting models to data}

\author[1]{Carlo M. Valensise}
\author[2,3]{Matteo Cinelli}
\author[2,*]{Walter Quattrociocchi}

\affil[1]{Enrico Fermi Research Center, Piazza del Viminale, 1 — 00184 Roma (IT)}
\affil[2]{Sapienza University of Rome — Department of Computer Science, Viale Regina Elena, 295 — 00161 Roma (IT)}
\affil[3]{Italian National Research Council — Institute for Complex Systems Via dei Taurini 19, 00185 Roma (IT)}

\affil[*]{Corresponding author: walter.quattrociocchi@uniroma1.it}
\date{}

\begin{document} 



\maketitle

\begin{abstract}
Users online tend to join polarized groups of like-minded peers around shared narratives, forming echo chambers. The echo chamber effect and opinion polarization may be driven by several factors including human biases in information consumption and personalized recommendations produced by feed algorithms. 
Until now, studies have mainly used opinion dynamic models to explore the mechanisms behind the emergence of polarization and echo chambers. The objective was to determine the key factors contributing to these phenomena and identify their interplay.
However, the validation of model predictions with empirical data still displays two main drawbacks: lack of systematicity and qualitative analysis.
In our work, we bridge this gap by providing a method to numerically compare the opinion distributions obtained from simulations with those measured on social media. To validate this procedure, we develop an opinion dynamic model that takes into account the interplay between human and algorithmic factors. We subject our model to empirical testing with data from diverse social media platforms and benchmark it against two state-of-the-art models.
To further enhance our understanding of social media platforms, we provide a synthetic description of their characteristics in terms of the model's parameter space. This representation has the potential to facilitate the refinement of feed algorithms, thus mitigating the detrimental effects of extreme polarization on online discourse.
\end{abstract}

\section*{Introduction}

The data-deluge~\cite{Bell2009} provided large amounts of data and digital traces of social activities that promised to improve the understanding of large-scale social phenomena~\cite{Lazer2009} exploiting the quantitative background ranging from mobility patterns \cite{barbosa2018human}, to information spreading \cite{Bakshy2015, DelVicario2016}.
Along this path, several studies focused on polarization dynamics \cite{bail2018exposure} especially about vaccines \cite{johnson2020online,schmidt2018polarization} or about the climate change debate \cite{dunlap2016political}.

Indeed, many human and environmental factors affect the spread and consumption of information online. 
On one side, human biases are the most relevant drivers for information selection, especially the confirmation bias~\cite{Nickerson1998, DelVicario2016}, such that people tend to privilege information aligned with their system of beliefs \cite{Bessi2015,Bakshy2015} and filter-out dissenting information \cite{Zollo2017}. Most of the online information spreads through social media platforms, whose business model aims to keep users as connected as possible; consequently, algorithmic choices may play a significant role in selecting which information is eventually proposed to the user \cite{balaji2021machine,zimmer2019fake}. The interplay between these factors is instrumental in shaping online social dynamics. One notable outcome is the emergence of echo chambers \cite{terren2021echo,centola_sciamerican,Cinelli2021}, groups of individuals who share similar views (particularly on contentious topics) against opposing perspectives \cite{DelVicario2016b}. 

Across echo-chambers, individuals opinions are polarized and such polarization may catalayze misinformation \cite{vicario2019polarization}, and burst other problematic phenomena such as segregation and hate speech \cite{cinelli2021dynamics}, that may eventually alter democratic processes~\cite{Sunstein2004}, and a cause for political dysfunction \cite{haidtbailongoing}.

At present, polarization metrics typically rely on aggregated activity data gathered from social media platforms~\cite{Cinelli2021, DeFrancisciMorales2021}. However, a more challenging task is detecting the underlying dynamics that give rise to a polarized network \cite{Waller2021}. Nevertheless, the onset of polarization in a network can be investigated through opinion dynamics models~\cite{Levin2021,Loreto2009}.
A recent review of opinion dynamics models is given in~\cite{Peralta2022}. Interestingly, the authors conclude that the main limitation of the modeling approach is the missing empirical validation. In a research field exposed to high volatility of scientific results~\cite{Ruths2019}, referencing and comparing models against observed behaviors is of crucial importance to correctly identify the drivers of these phenomena.

In this paper, we address this issue studying a flexible opinion dynamic model capable of reproducing a wide range of opinion distributions over a social network, depending upon the strength of algorithmic bias. Next, we employ the Jensen-Shannon (JS) divergence to quantitatively evaluate the degree of conformity between our simulations and empirical data collected from four social media platforms. By grid-searching model's parameter space we are able to identify the most suited configurations to describe each platform. Importantly the JS-based procedure can be applied to potentially any model whose outcome provides an opinion distribution over a network. In fact, we compare the capability of our model with that of two recent models~\cite{Baumann2020,arruda2022}. 

Our research offers significant contributions to the field of social dynamics modeling. Firstly, we propose a new opinion dynamics model that considers human attitudes and algorithmic features in information consumption and propagation processes. 
Our model outperforms others in capturing the dynamics of extreme polarization, as evidenced by our procedure for computing model-data agreement. Secondly, we provide a quantitative approach for assessing the degree of conformity between model outcomes and social media data. This procedure can be applied to any opinion dynamics model, enabling direct comparison between different models and empirical data. Moreover, it provides a parametric description of the opinion state on a social media platform, which could be employed to fine-tune recommendation algorithms and mitigate extreme polarization.

\section*{Related Work}

Opinion dynamics models are rooted in traditional agent based models whose aim is reproducing a wide set of social dynamics ranging from residential segregation~\cite{schelling1971dynamic} to cultural dissemination~\cite{axelrod1997dissemination}.
In opinion dynamics models, agents are usually represented as nodes of a graph endowed with some properties, namely opinions or attitudes~\cite{Loreto2009,Perc2013,Baronchelli2018}. Connections among nodes may represent social relationships (e.g. friendship) and allow agents to interact with each other. 
Simulations consist of updating agents' internal states and/or network connections (rewiring) based on the opinions of neighboring others. The opinion update depends upon the specific model that may involve (i) the fundamental mechanism of homophily, by which individuals tend to interact and create connections with others sharing similar features~\cite{McPherson2001,Lee2019,blex2022positive} (ii) social contagion~\cite{Aral2009, Centola2007}, i.e. the tendency of individuals to become similar with each-other over time; (iii) algorithmic bias, namely the content selection mechanism of online social media platforms~\cite{Perra2019,Santos2021,Bonchi2020}. 

One of the most used frameworks for modeling opinions in a continuous space is the Bounded Confidence Model (BCM)~\cite{deffuant2001mixing} in which the opinion update of an agent is ruled by a tolerance parameter imposed on the difference between the opinion value of the agent and the neighbors. Considering the BCM as a basis, a few variants of that model simulated the emergence of polarization in the opinion space considering different factors such as the role of traditional and new media~\cite{Quattrociocchi2014} and biases in information consumption~\cite{del2017modeling}.

As reported in~\cite{Peralta2022}, the set of opinion dynamics models is vast and the main challenges consist in using model outcomes beyond providing proofs of concept and eventually comparing the performance of different models in reproducing what observed in the data. For these reasons, in this section, we review two state-of-the-art models by Bauman et al.~\cite{Baumann2020} and de Arruda et al.~\cite{arruda2022} that will be used in a comparative analysis provided in the Results section.

The model proposed by Baumann et al.~\cite{Baumann2020} considers a dynamic network in which the opinions dynamic is driven by interactions among agents, described by a system of coupled ordinary differential equations
\begin{equation}
\dot{x}_i = -x_i + K\sum_{j=1}^{N} A_{ij}\,\tanh(\alpha\,x_j)\,,
	\label{eqn:starnini}
\end{equation}
where $x_i$ is the opinion of $i-th$ user, $A$ is the adjacency matrix of the network, and $K$, $\alpha$ are free parameters of the model respectively representing the social interaction strength among agents and the \textit{controversialness} of the modeled topic. The social network underlying the opinion dynamics is updated at every iteration, with the probability of two users being connected proportional to 
\[
p_{ij}\sim \lvert x_i-x_j \rvert^{-\beta}\,.
\]
The last free parameter is the reciprocity $r$, i.e. the probability that given a established link $i \to j$, it is reciprocated from $j \to i$.

In the model proposed by de Arruda et al.~\cite{arruda2022} the simulation is performed at the level of users producing information (posts) expressing a given opinion $\theta$, and reaching neighboring users. In this model, interactions between individuals occur probabilistically, resulting in neighbors updating their opinions to become closer or farther from the opinion expressed in the post. Various outcomes can be achieved by adjusting the post transmission probability, post distribution probability, relative phase parameter, and the ability to rewire connections. The transmission probability distribution, which gives the probability for a user to post a piece of information, as a function of $d=\lvert x_i-\theta \rvert$, being $x_i$ the opinion of user $i$, and may assume three shapes:
\[
P_{t}^{uni} = 1\qquad P_t^{pol} =  \cos^2\left(d\,\pi/2\right)\qquad P_t^{sim} = \begin{cases}
	\cos^2\left(x \pi/2\right)\, \text{if}\, x\le 1\\ 0\,.
\end{cases}
\]
The $uni$ distribution models the case in which the users post content without caring about the post opinion $\theta$, thus they spread everything they come across. The $pol$ distribution models the case in which the users tend to post content that can be either very similar or very diverse with respect to their opinions. Finally, the $sim$ case models the scenario in which users post like-minded content with a much higher probability.

A similar mechanism is considered for neighboring users that receive the information depending on their opinion distance from user $i$ (post distribution), according to two possible probability distributions. 
\[
P_I(y)= \cos^2\left(y\frac \pi 2+\phi\right)\qquad P_{II}(y)= \cos^2\left(\frac y2 \frac \pi 2+\phi\right) \,,
\]
being $y=\lvert x_j-x_i\rvert$ and the phase $\phi$ a free parameter. 
In other words, the post transmission refers to the probability of users posting a certain post on their timeline, this post is then transmitted to their neighbors according to the filtering imposed by the recommendation algorithm which is modeled by the post distribution probability.
The model investigates also the role of rewiring, which, if allowed, takes place with a probability proportional to $y$.

\section*{Methods}

\subsection*{Opinion dynamics model}

\subsubsection*{Model description}

\begin{figure}[htbp]
	\centering
	\includegraphics{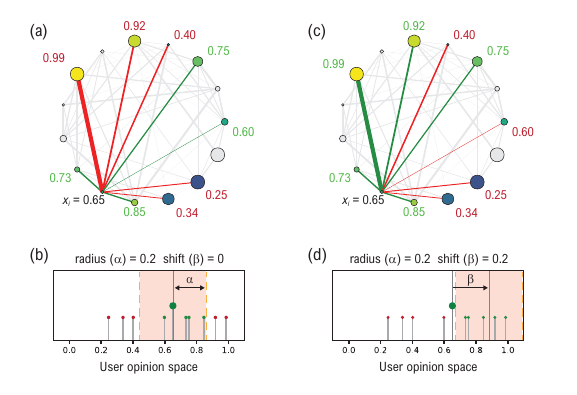}
	\caption{Opinion dynamics model with algorithmic bias. We consider an Erd\H{o}s-Rényi network; each node represents a user with a given firing rate ($\sigma_i$, encoded by size) and opinion ($x_i$, color); weighted edges represent the strength of relationships, green links contribute to the update of opinion $x_i$ and get reinforced by a $\gamma$ factor, red links are weakened by the same factor, to represent the goal of platforms to comply with users preferences; (a-b) the interaction radius $\alpha$ determines the actual neighborhood seen by the user; (c-d) effect of shift $\beta\ne0$ on the neighborhood.}
	\label{fig:figure1}
\end{figure}

We consider a social network in which each node represents a user having an opinion $x_i$ with a firing rate $\sigma_i$ determining $i$'s activation, and in which edges are weighted friendship relations (see panels a and c of Figure~\ref{fig:figure1}). During an iteration, the opinion of each user is updated according to a filtered subset of its nearest neighbors. The filtering mechanism mimicks the feed/recommendation algorithm's action on social media platforms that play a major role in shaping users online experience~\cite{huszar2022algorithmic}. 

Algorithmic and cognitive bias effect are ruled by three parameters: radius $\alpha$, shift $\beta$ and discount $\gamma$. 
At every time step, users undergo an opinion update, after interacting with their friends. We associate every time step to a session on a social media in which the feed algorithm proposes contents to users. We assume that platforms aim at maximizing the permanence of users, thus proposing contents aligned with each users' viewpoint~\cite{sunstein2018}. To this aim, the radius $\alpha$ acts as a boundary for the opinion space accessible to a user. It sets the maximum opinion distance reachable by a user, thus forcing the opinion to be updated according to like-minded connections. The shift $\beta$ instead, models the tendency to drive opinion towards extreme values, as it corresponds to a shift of the accessible opinion interval. The last parameter $\gamma$ models the tendency to consolidate explored connections and weaken links with nodes excluded from the opinion interaction interval. Moreover, we consider two more parameters $a$ and $b$ ruling the initial opinion distribution. As illustrated in Figure~\ref{fig:figure1}(b-d), a user with opinion $x_i$ interacts with neighboring users whose opinion falls in the interval $[x_i -\alpha + \beta, x_i+\alpha+\beta]$. The connections with these users are reinforced by a factor $\gamma$, while relations with excluded users are weakened by the same factor.

\subsubsection*{Model implementation}

{The users are embedded in a social network generated using the Erd{\H{o}}s--R{\'e}nyi model with $N=5000$ nodes and mean degree $\langle d\rangle=6$. Edges are randomly weighted, such that $w_{ij}\sim\mathcal U(0,1)$ represents the strength of social relationships between user $i$ and $j$.} Pictorial representations are displayed in Figure~\ref{fig:figure1}(a-c).
Each node $i$ has an activation (or firing) rate, randomly sampled from a uniform distribution $\sigma_i \sim \mathcal U(0,1)$ and opinion $x_i\sim \mathcal P(x)$. 
{Assuming $\mathcal P(x)$ uniform in the $[0,1]$ interval,} this random initialization corresponds to the leaning distribution shown in Figure~\ref{fig:figure1}(b), in which neither opinion segregation nor polarization is present. 

At each discrete time-step, sequentially, each node randomly undergoes an interaction, with probability $\sigma_i$, represented as node size in Figure~\ref{fig:figure1}. If user $i$ is active, then its opinion gets updated according to the opinion of its first neighbors $\mathcal N(i)$ and the effect of platform filtering, similarly to other bounded confidence implementations~\cite{Quattrociocchi2014}.

In absence of algorithmic bias, the opinion of the selected user $x_i$ is updated as
\[
x_i^{t+1} = x_i^t +\delta\,,
\]
with 
\[
\delta =  \mu\, \left( x_{\mathcal N(i)}^t- x_i^{t}\right) \,,
\]
being $x_{\mathcal N(i)}^t$ the average weighted neighborhood opinion
\begin{equation*}
x_{\mathcal N(i)}^t =  \dfrac{\sum\limits_{j \in \mathcal N(i)}^{}\, x_j^t \,w_{ij}}{\sum\limits_{j \in \mathcal N(i)}\,w_{ij}} \,,  
\end{equation*}

where $\mu$ is a convergence factor, set to $\mu = 0.3$. 

Algorithmic and cognitive biases are introduced in the model as filters on the neighborhood $\mathcal N(i)$ of each user, which depend on three parameters: interaction radius $\alpha$, opinion shift $\beta$ and discount $\gamma$. With reference to Figure~\ref{fig:figure1}{(b-d)}, the algorithm selects neighbors whose opinion lies in the interval $[x_i-\alpha+\beta, x_i+\alpha+\beta]$. The sign of the opinion shift $\beta$ is given by 
\[
\operatorname{sign}(\beta) = \operatorname{sign}(x_i^t -0.5)\,.
\]
These parameters govern the actual interaction of users in shaping the opinion dynamics. Only {suitable neighbors (i.e. neighbors falling within the interval)} effectively contribute to the update of user opinion, such that
\[
x^t_{\mathcal N(i)} = x^t_{\mathcal N(i,\alpha, \beta)}\,.
\]
After the update of user opinion, links within the filtered neighborhood are reinforced, while those outside are discounted, depending on the value of $\gamma$: 
\[
w_{ij}^{t+1} =
\begin{cases}
	w_{ij}^t(1+\gamma) \quad \text{if}\, j \in \mathcal N(i,\alpha,\beta)\\
	w_{ij}^t(1-\gamma) \quad \text{if}\, j \notin \mathcal N(i,\alpha,\beta)\\
\end{cases}
\]
Opinion and weight values are forced in the $[0,1]$ interval through clipping. {The number of iterations is fixed to $T = 500$ however,} both the network and opinion dynamics generally converge to a stable state after 50-100 time steps (see Figure S1). The final distribution of opinions is determined by the evolution of user opinions and network connections over time.

Finally, the steps of the model can be resumed as follows:
\begin{enumerate}
    \item Activation of user $i$ with probability $\sigma_i$
    \item Visualization of the suitable neighbors in the window $[x_i -\alpha + \beta, x_i+\alpha+\beta]$
    \item User opinion update $x_i^{t+1} = x_i^t +\delta$
    \item Links strength update by a $\gamma$ factor
\end{enumerate}

\subsection*{Fit procedure}
Beyond the three parameters ($\alpha$, $\beta$, $\gamma$) representing the effect of algorithmic bias, the model evolution depends upon the initial opinion distribution $P(x)$ from which users opinions are sampled. We consider the beta distribution, 
\[
P(x; a, b) \propto x^{a-1}\,(1-x)^{b-1}\,,
\]
ruled by $a>0$ and $b>0$ with support in the $[0,1]$ range. Thus, the free parameters set of the model is $\sigma = \{\alpha, \beta,\gamma,a, b\}$.

To characterize the model outcomes and compare them with data measured on social media, we performed a grid search {(i.e., a near-exhaustive search over a specified set of parameters)} of 900 different parameter configurations. For each parameter the variation range is expressed in Table~\ref{tab:table1}.
\begin{table}[htbp]
	\centering
	\begin{tabular}{cr}
		\hline
		Parameter & Values (start, stop, step) \\
		\hline
		$\alpha$ & (0.1, 1, 0.1) \\
		$\beta$ & (0, 1, 0.1)\\
		$\gamma$ & 10\textsuperscript{$\wedge$}(-3, -1, 1)\\
		$a$ & (1,3,1) \\
		$b$ & (1,3,1)\\
		\hline
	\end{tabular}
	\caption{Parameter variability range}
	\label{tab:table1}
\end{table}

For each final opinion configuration, we computed the kernel-density estimate for the joint probability distribution of users and neighborhood opinions. This allows us to directly compare model outcomes with measured data~\cite{Cinelli2021}. In more detail, we discretized the opinion space into a grid to compare the opinion distribution generated by our model with the empirical data obtained from social media platforms and we collapsed it into an ordered vector of discrete symbols. Then, using an appropriate similarity measure, we quantified the agreement between the model and data opinion vectors.

The agreement between a model realization and data is quantitatively evaluated by the Jensen–Shannon (JS) divergence, that is the symmetric version of the Kullback-Leibler (KL) divergence between two distributions $P$ and $Q$
\[
D(P\parallel Q) = \int_{-\infty}^{+\infty} P(x)\,\log\left(\frac {P(x)}{Q(x)}\right)\,,
\]
where $P$ and $Q$ are probability densities. {$P$ usually represents a distribution obtained from the data while $Q$ represents a model or a distribution used for comparison. It follows that the $KL$ divergence can be interpreted as the average distance (in number of bits) required for encoding samples from $P$ using a code optimized for the distribution $Q$ rather than for $P$. In other words, the KL divergence can be interpreted as the expectation of the logarithmic difference between the probability distributions $P$ and $Q$ when the expectation is computed using the distribution $P$.}
From this definition the JS divergence is defined as
\[
\begin{split}
	\operatorname{JSD}(P\parallel Q) &= \frac 12 D (P\parallel M) + \frac 12 D(Q\parallel M)\,\,,\\
	&\text{with}\quad M=\frac 12 (P+Q)\,.
\end{split}
\]
{The JS divergence quantifies the difference between two probability distributions $P$ and $Q$ by computing the KL divergence in both directions with respect to a synthetic distribution $M$ obtained averaging the two distributions used as inputs. The JSD varies in range [0,1] when using the base 2 logarithm; low values of the measure indicate similarity between the distributions $P$ and $Q$ while high values indicate, in fact, divergence.}
The same procedure is applied for the alternative models considered.


\section*{Results}

\begin{figure}[htbp]
	\centering
	\includegraphics{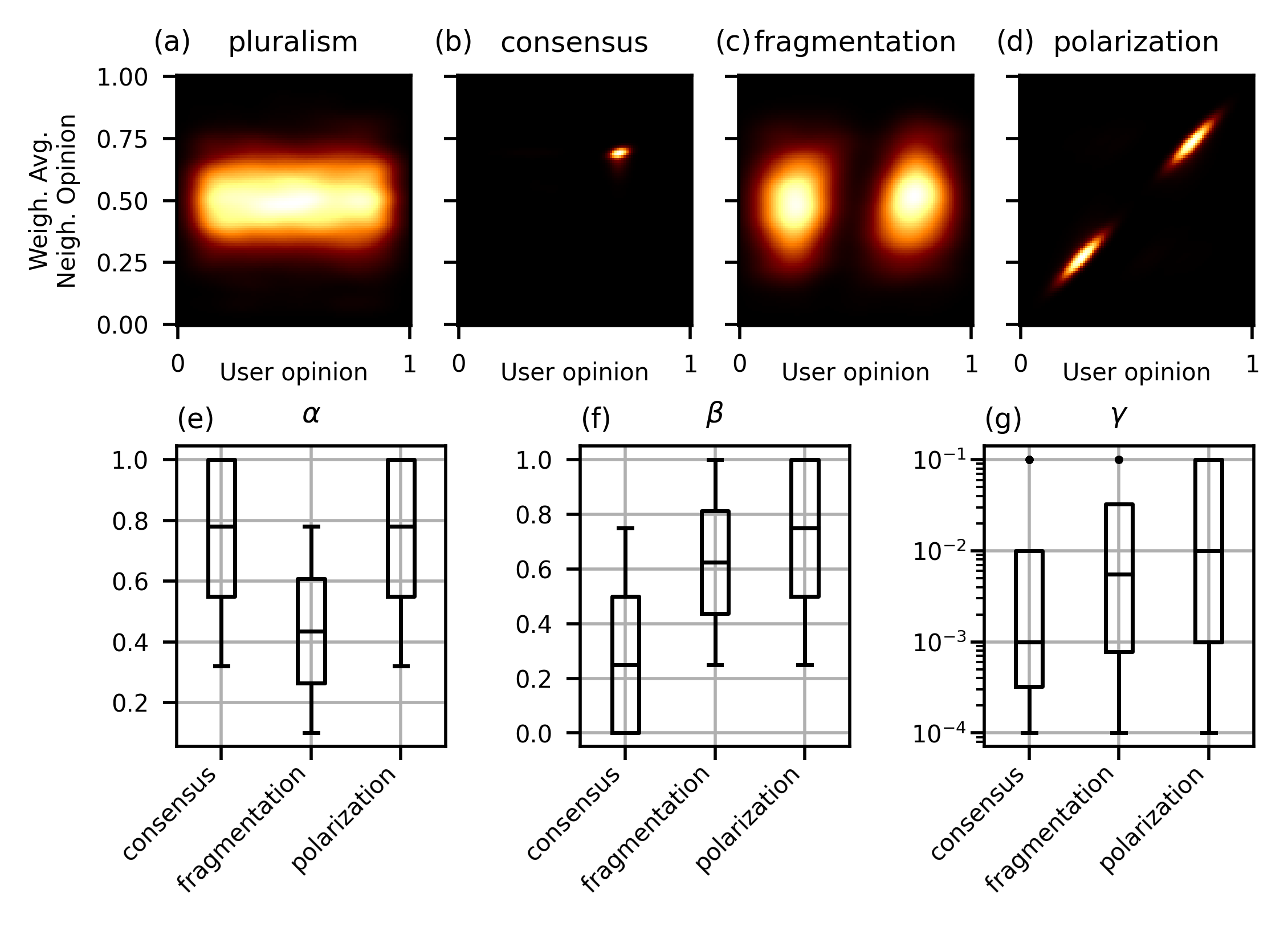}
	\caption{(a) Initial opinion distribution. (b-d) Exemplary model outcomes for opinion distribution representing consensus, fragmentation and polarization of opinion space. (e-g) Boxplots of $(\alpha, \beta, \gamma)$ for each kind of opinion distribution.}
	\label{fig:figure2}
\end{figure}

The model idea roots in the Bounded Confidence Model~\cite{deffuant2001mixing}, with a paradigmatic shift from the user-centric dynamic to an algorithmic-centric one. In fact, with respect to BCM we do not consider the intrinsic limitation of user interactions with respect to opinion distance; we focus on the role of algorithmic bias that prevents interactions among users with distant opinions, with a tunable level of severity. 
Depending upon the parameters set $\sigma$, the model evolution produces different kind of opinion distributions on the network $\delta(\sigma)$, computed as the kernel-density estimate for the joint probability distribution of users and neighborhood opinions. 
In Figure~\ref{fig:figure2}(a-d) we report examples of characteristic distributions in the opinion space. Pluralism is a state in which all opinions are represented in the network and each user is surrounded by a multitude of opinions; consensus~\cite{Baronchelli2018} is a condition in which all network opinions collapse into a very narrow region. Conversely, fragmentation~\cite{Srbu2019} is very similar to pluralism, apart from the fact that not all the opinion space is available. Eventually, polarization~\cite{garret09,Cota2019,DelVicario2016} is described as a fragmented and segregated opinion space: two main opinions are present, and users expressing each of them are connected with like-minded others.

To investigate the role of $(\alpha, \beta, \gamma)$, starting from a pluralism distribution, we computed the evolution for 900 different parameter combinations $\mathcal S=\{\sigma_1,\dots  \sigma_{900}\}$ (see Methods). Most of the produced opinion distributions can be classified in the above-mentioned categories. Results are reported in Figure~\ref{fig:figure2}(e-g). As expected, a configuration in which users have the possibility to interact with many others while having low shift and discount leads to consensus. Low $\alpha$ values, with moderate $\beta$ and $\gamma$, lead to fragmentation, mainly due to the narrow opinion space accessible by users. Conversely, the parameter $\gamma$, that rules the forget/reinforce mechanism, appears to be a pivotal driver to reach a polarized opinion space in which echo chambers arise.

\begin{figure}[htbp]
	\centering
	\includegraphics{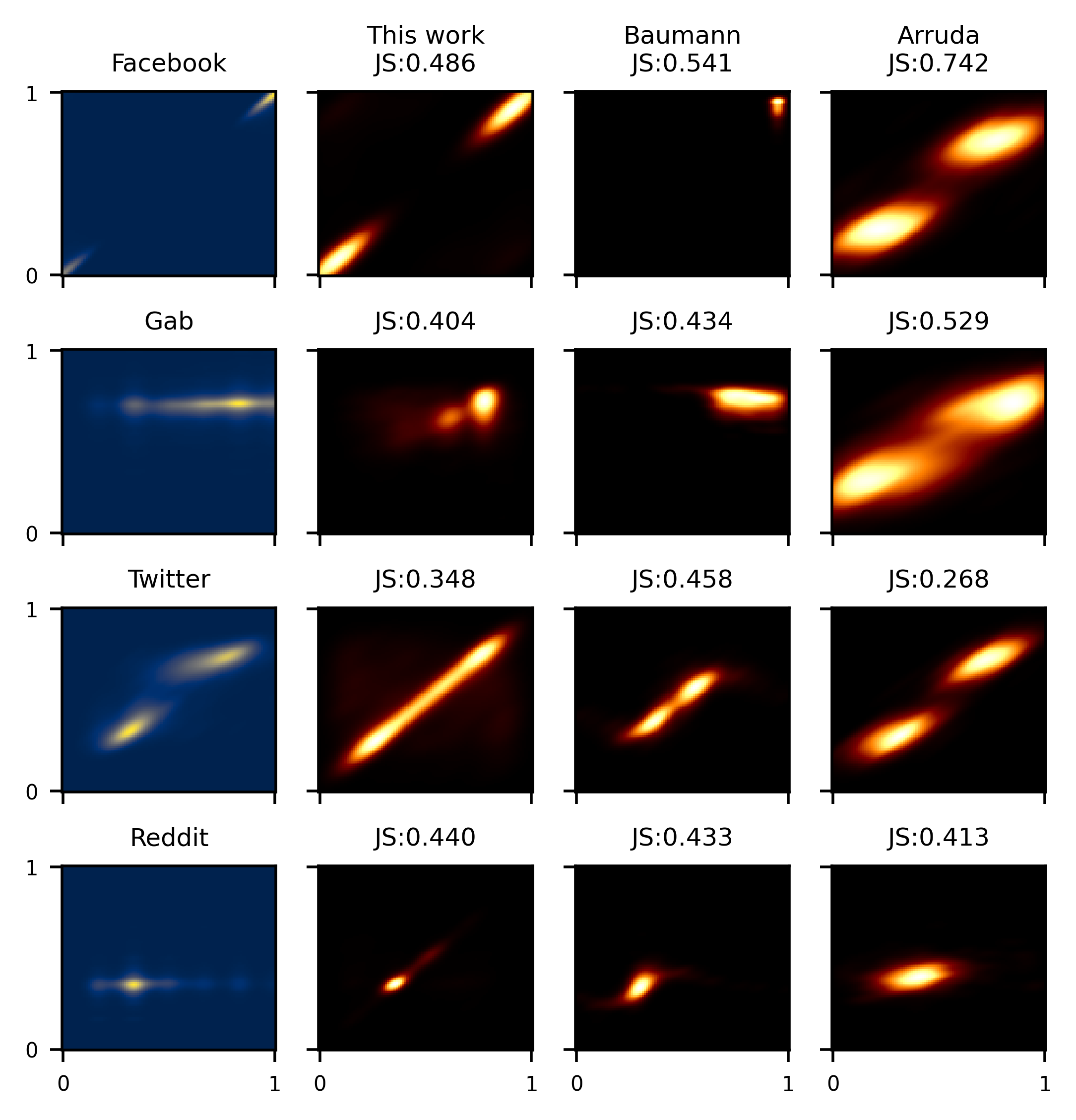}
	\caption{Evaluation of model outcomes adherence to real data. Left column: real opinion distributions measured in \cite{Cinelli2021}. From second to last column: best approximation of the observed data provided by our and competing opinion dynamic models: Baumann~\cite{Baumann2020} and de~Arruda~\cite{arruda2022}.}
	\label{fig:figure3}
\end{figure}

Despite the large number of opinion dynamics models present in literature, a crucial missing point is a quantitative comparison with real opinion data from social media platforms. To address this point, we consider four examples of opinion distributions $\mathcal P=\{\pi_F, \pi_T, \pi_G, \pi_R\}$ measured on four different social media platforms, namely Facebook, Twitter, Gab and Reddit, using the same data of~\cite{Cinelli2021}. Specifically, Facebook data relates to users interactions on discussions about vaccines; for Twitter, data is made up of discussions about abortion; Reddit and Gab datasets contain interactions of users with political or news contents. {Further details on the employed datasets and on the data collection are reported in Supplementary Information and in reference~\cite{Cinelli2021}.}
Next, we quantitatively evaluate the difference of each empirical opinion distribution $\pi_p$ with all the opinion distributions computed via grid-search $\mathcal D=\{\delta (\sigma_i)\,, \sigma_i \in \mathcal S \}$, by computing the Jensen-Shannon divergence (JSD), a quantity suitable to compare distributions. For each platform $p$ the best parameters set $\sigma^*_p$ is identified as 
\[
\sigma^*_p = \underset{\sigma \in \mathcal S}{\arg\min}\,\operatorname{JSD}\left(\delta(\sigma), \pi_p\right)\,,
\]
namely considering the minimum JSD. {In more detail, for each of the four social networks we obtained a set of 900 JSD values (one per each parameter combination of the grid search) of which extract the minimum.}
The best approximating distributions are shown in Figure~\ref{fig:figure3}, along with the optimal distribution obtained applying the same procedure to two recent opinion dynamics models, from Baumann et al.~\cite{Baumann2020} and de~Arruda et al.~\cite{arruda2022}. 

All models show the flexibility required to well approximate various scenarios. Interestingly, no single model can well approximate all platform-related opinion distributions (see Discussion).

Importantly, we observe that all three models display degeneracy with respect to the optimal solution for each platform. In fact, multiple configurations produce values of JSD very close each other, as shown in Supplementary Information (Figure S2). Thus, a more comprehensive description of each social media, in terms of model parameters, can be obtained by looking at all the configurations whose corresponding JSD falls within the 5\textsuperscript{th} percentile of the JSD distribution. In other words, for each platform, we select the 5\% of observations showing the lowest values of JSD, that is, a sufficiently high similarity with the data.

\begin{figure}[htbp]
	\centering
	\includegraphics{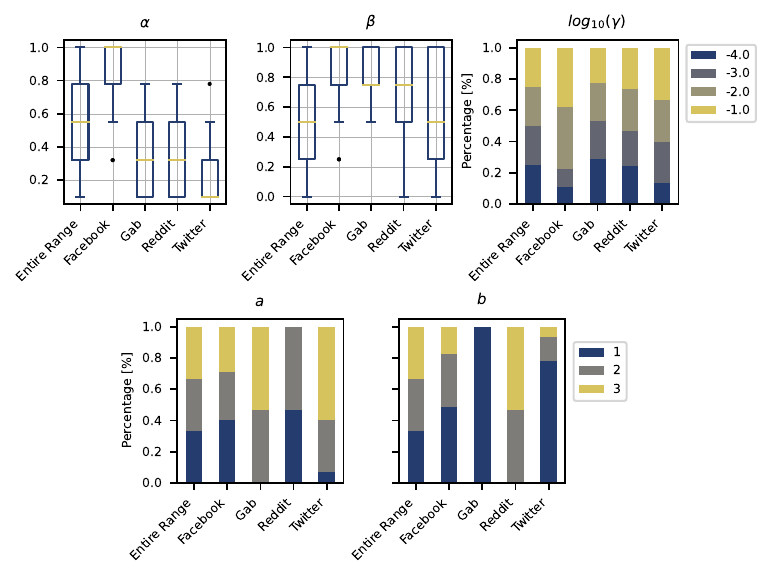}
	\caption{For each social media, we consider the top 5\% approximating configurations, and display the corresponding parameter distribution in each panel, together with the full parameter range explored with the grid-search. This visualization allow to interpret the relevant range for each parameter and each social media. Stacked bar charts represent what percentage of the top 5\% of the configurations with the smallest JSD have the parameter values displayed in the legend. The parameter $\alpha$ is the radius of opinions observable by the user, $\beta$ is the shift imposed to the radius, $\gamma$ is the discount factor used for the correction of links weight while $a$ and $b$ are the parameters of the beta distribution used for modeling the initial opinion distribution.} 
	\label{fig:my_mode_distro}
\end{figure}

In the case of our model, the results are reported in Figure~\ref{fig:my_mode_distro}. Each panel allows to asses the relevant values of each parameter in approximating the distribution from each social network, with respect to the entire grid-searched range. The best approximation for Facebook is obtained with relatively high values of $\alpha$, $\beta$ and $\gamma$, corresponding to a relevant presence of algorithmic filtering. On Twitter, the leading parameter is the small interaction radius $\alpha$, with a relevant dependency on the initial opinion distribution. Bias and reinforcement/discount of network connection play a less relevant role. This observation could highlight differences in the algorithmic filtering between the two platforms. On Reddit and Gab the opinion space related to news and politics discussions does not show polarization. The best fit of our model to these configurations corresponds to a mild algorithmic effect, expressed by the non-trivial combination of $\alpha$ and $\beta$; the most relevant role is instead exerted by $a$ and $b$, shaping the initial opinion distributions.

\begin{figure}[htbp]
	\centering
	\includegraphics{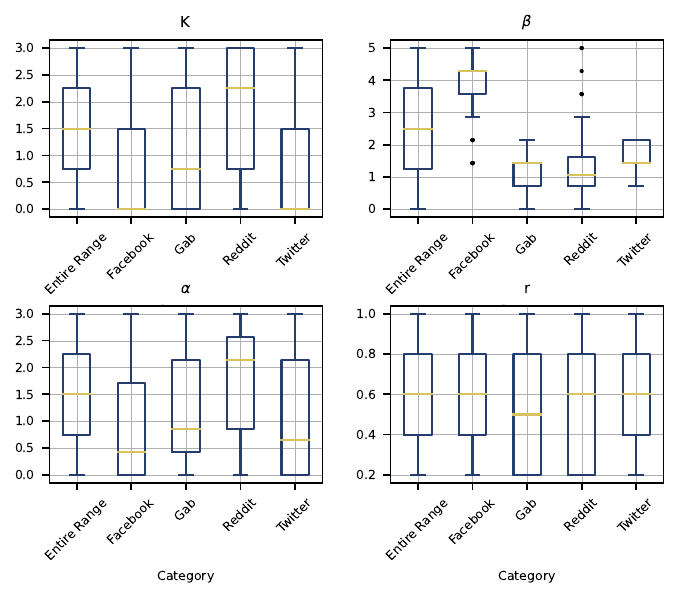}
	\caption{Fit of Baumann et al.~\cite{Baumann2020}. For each social media, we consider the top 5\% approximating configurations, and display the corresponding parameter distribution in each panel, together with the full parameter range explored with the grid-search. This visualization allow to interpret the relevant range for each parameter and each social media. 
 Stacked bar charts represent what percentage of the top 5\% of the configurations with the smallest JSD have the parameter values displayed in the legend. The parameter $K$ models the strength of the social interaction; $\beta$ is the power-law exponent used for computing the probability of connection between two users; $\alpha$ models the \textit{controversialness} of the topic; and $r$ is the reciprocity in establishing social ties.
 } 
	\label{fig:star_distros}
\end{figure}

\begin{figure}[htbp]
	\centering
	\includegraphics{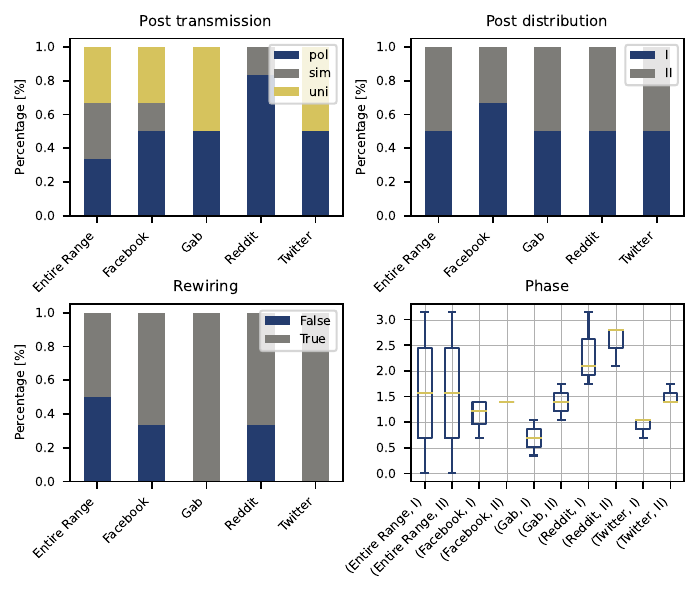}
	\caption{Fit of de~Arruda et al.~\cite{arruda2022}. For each social media, we consider the top 5\% approximating configurations, and display the corresponding parameter distribution in each panel, together with the full parameter range explored with the grid-search. This visualization allow to interpret the relevant range for each parameter and each social media. {Stacked bar charts represent what percentage of the top 5\% of the configurations with the smallest JSD have the parameter values displayed in the legend. The post transmission refers to the probability of users posting a certain post on their timeline, this post is then transmitted to their neighbors according to the filtering imposed by the recommendation algorithm which is modeled by the post distribution probability. In the post transmission panel: the $pol$ case users post only contents at the extreme of the opinion space; in the $sim$ case the users post only like-minded content; in the $uni$ case the users post every type of content. In the post distribution panel: I and II refer to two different transmission probability functions, the smoother of which in the second case. In the rewiring panel: False and True refer to the presence or absence of rewiring in the modeling process.}
	} 
	\label{fig:moreno_distros}
\end{figure}

The same analysis was conducted for the two other opinion dynamics models considered. Results are reported in Figure~\ref{fig:star_distros} and Figure~\ref{fig:moreno_distros}. Despite the rather different working principle, the model by Bauman et al.~\cite{Baumann2020}, displays good performances, except for not capturing the extremely polarized opinion distribution observed on Facebook. This results show that the main driver to approximate different opinion distributions is given by the power law exponent $\beta$, and the \textit{controversialness} $\alpha$, with a less relevant role played by $K$ and $r$. 

From Figure~\ref{fig:moreno_distros} we observe that, for the social where the model by de~Arruda et al.~\cite{arruda2022} performs best, namely Reddit and Twitter, the kind of post distribution function play a less relevant role; the phase distributions $\phi$ in the top 5\% of configurations instead show much more localized values, suggesting a relevant role of this parameter. With respect to post transmission and rewiring, in Reddit we observe a prevalence of the \textit{pol} {(users post only contents at the extreme of the opinion space)} type, and a balanced representation of \textit{uni} {(users post every type of content)} and \textit{sim} {(users post only like-minded content)}. Lastly, the possibility to rewire connection is key to well approximate the opinion distributions.

\section*{Discussion}

Polarization is a challenging and multifaceted phenomenon with psychological, social, and policymaking implications. It can be viewed as both a state and a process~\cite{Waller2021}, resulting from the interplay between opinion distribution and network structure. Thus, understanding the social and algorithmic mechanisms that drive online polarization is crucial in addressing its potential negative effects.

Collective alignment in social networks can be gauged by pooling interaction data, as elucidated by recent studies~\cite{Cinelli2021,DeFrancisciMorales2021}. Such \textit{a posteriori} measurements of polarization serve as a benchmark for developing models that can effectively replicate them. Such an approach enables to map opinion distributions, stemming from interaction networks, onto a specific parameter set of a given model. With the attainment of satisfactory agreement between models prediction and empirical data, the model itself could be employed as a starting point for performing scenario analysis.

Moreover, by comparing the performances of our model with others, we demonstrate that a \textit{silver-bullet}, in terms of opinion dynamics models, i.e. a model able to capture the full spectrum of real-world opinion distributions does not yet exist. Notably, however, the models do perform well regardless of the level at which the data are collected: with respect to Gab and Reddit we are observing opinion distributions sampled at the community level, while for Twitter and Facebook the data are specifically related to a given topic (abortion and vaccination, respectively). Including more complex mechanisms ruling the opinion dynamics may lead to better numerical results at the cost of preventing a simpler interpretation.

\section*{Conclusion}
In this work we propose a novel, flexible opinion dynamics model, inspired by a set of quantitative studies, that considers the algorithmic dimension of online interactions among users; second, we ``close the loop'' by quantitatively measuring the adherence of model outcomes to empirical data, and obtain a description of social media platforms in terms of model parameters. This aspect is particularly relevant in presence of extreme polarization, a phenomenon associated to segregation of users in echo chambers, that in turn could foster problematic phenomena (e.g. misinformation spreading). 
Our study has yielded a noteworthy finding: the parameter $\gamma$, which governs the algorithmic strengthening and weakening of the relative influence of neighbors, plays a crucial role in reducing polarization, as predicted by our model. In other words, by manipulating $\gamma$, we can effectively control the level of polarization. The dynamics of opinion formation and social influence are complex phenomena that require a multi-faceted approach to comprehend fully. Scholars must delve into not only the shifting beliefs of individuals but also the underlying network structure that gives rise to them. Investigating these structural aspects as both a starting point and an evolving influence in opinion formation processes is crucial for understanding the dynamics at play.
Moreover, given the present state of our society, there is an urgent need to explore how we might disrupt the polarized state of social systems. The study of what actions and parameter values are necessary to achieve this end is thus an important area for future inquiry.
By delving deeper into the mechanisms of opinion polarization and the potential strategies that could be employed to mitigate its effects, researchers can provide valuable insights to policymakers and stakeholders alike.



\section*{Acknowledgments}
We thank Michele Starnini for the fruitful discussions. We also thank Henrique Ferraz de Arruda for providing the code to reproduce the model.
The work is supported by IRIS Infodemic Coalition (UK government, grant no. SCH-00001-3391), 
SERICS (PE00000014) under the NRRP MUR program funded by the European Union - NextGenerationEU, project CRESP from the Italian Ministry of Health under the program CCM 2022, and PON project “Ricerca e Innovazione” 2014-2020 and project SEED n. SP122184858BEDB3.
 
\section*{Supplementary information}

SI file



\end{document}